\documentclass[twocolumn]{aastex631}

\usepackage{amsmath, amssymb, amstext}
\usepackage{graphicx}
\usepackage{natbib}
\usepackage[ruled,vlined]{algorithm2e}
\usepackage{appendix}

\newcommand{\bx}{x}
\newcommand{\be}{e}

\newcommand{\E}{\mathrm{E}}
\newcommand{\msun}{$M_{\odot}$}
\newcommand{\mzams}{{M_{\text{ZAMS}}}}
\newcommand{\mbh}{{M_{\text{BH}}}}
\newcommand{\mns}{{M_{\text{NS}}}}
\newcommand{\bybh}{{\tilde{M}_{\text{BH}}}}
\newcommand{\bzbh}{{\mbh}}
\newcommand{\byns}{{{\tilde{M}_{\text{NS}}}}}
\newcommand{\bzns}{{\mns}}
\newcommand{\by}{{\bybh}}
\newcommand{\bz}{{\bzbh}}
\SetKwInput{KwDefine}{Notation}
\SetKwInput{KwInput}{Input}
\SetKwInput{KwOutput}{Output}

\shorttitle{
Statistics, Supernovae, and the Nuclear Equation of State
}
\shortauthors{Meskhi et al.}

\turnoffeditone  % this turns off the edits from round 1
\turnoffedittwo  % this turns off the edits from round 2

\graphicspath{{./}{figures/}}

\begin{document}

\title{
A New Constraint on the Nuclear Equation of State from Statistical Distributions \\ of Compact Remnants of Supernovae
}

\correspondingauthor{Carla Fr\"ohlich}
\email{cfrohli@ncsu.edu}

\author[0000-0001-9152-3955]{Mikhail M. Meskhi}
\affiliation{
    Department of Computer Science, 
    University of Houston, 
    Houston TX 77204-3010, USA 
}

\author[0000-0003-2540-3845]{Noah E. Wolfe}
\affiliation{
    Department of Physics, 
    North Carolina State University, 
    Raleigh NC 27695 USA
}

\author[0000-0001-6135-7749]{Zhenyu Dai}
\affiliation{
    Department of Computer Science, 
    University of Houston, 
    Houston TX 77204-3010, USA 
}

\author[0000-0003-0191-2477]{Carla Fr\"ohlich}
\affiliation{
    Department of Physics, 
    North Carolina State University, 
    Raleigh NC 27695 USA
}

\author[0000-0001-6432-7860]{Jonah M. Miller}
\affiliation{
    CCS-2, Computational Physics and Methods, 
    Los Alamos National Laboratory, 
    Los Alamos NM 87544 USA
}
\affiliation{
    Center for Theoretical Astrophysics, 
    Los Alamos National Laandoratory, 
    Los Alamos NM 87544
}

\author[0000-0001-9342-3755]{Raymond K. W. Wong}
\affiliation{
    Department of Statistics, 
    Texas A\&M University, 
    College Station TX 77843 USA
}

\author[0000-0001-8165-8805]{Ricardo Vilalta}
\affiliation{
    Department of Computer Science, 
    University of Houston, 
    Houston TX 77204-3010, USA 
}

%% word limit: 250
\begin{abstract}
Understanding how matter behaves at the highest densities and temperatures is a major open problem in both nuclear physics and relativistic astrophysics. Our understanding of such behavior is often encapsulated in the so-called high-temperature nuclear equation of state, which influences compact binary mergers, core-collapse supernovae, and other phenomena. Our focus is on the type (either black hole or neutron star) and mass of the remnant of the core collapse of a massive star. For each of six candidate equations of state, we use a very large suite of spherically symmetric supernova models to generate a sample of synthetic populations of such remnants. We then compare these synthetic populations to the observed remnant population. Our study provides a novel constraint on the high-temperature nuclear equation of state and describes which EOS candidates are more or less favored by an information-theoretic metric.
\end{abstract}

%%%%%%%%%%%%%%%%%%%%%%%%%%%%%%%%%%%%%%%%%%%%%%%%
\section{Introduction} \label{sec:intro}
%%%%%%%%%%%%%%%%%%%%%%%%%%%%%%%%%%%%%%%%%%%%%%%%

Neutron stars (NSs) are some of the densest observable objects in the Universe. They provide a laboratory to probe matter at the highest densities -- a major open problem in astrophysics and nuclear physics. 
Understanding how matter behaves at extreme conditions, i.e.\ the high-temperature nuclear equation of state (EOS), is needed to predict and analyze the gravitational wave signals from NS mergers, the associated kilonova, pulsar properties, the outcome of core-collapse supernovae (CCSNe), and many more. 

Although the fundamental interactions are well-understood, the full many-body quantum-mechanics problem is currently intractable.
Moreover, 
this state of matter is not directly accessible through terrestrial experiments
\citep{Lonardoni2020}.

The traditional way of constraining the EOS is to examine the relationship between the mass and radius of a NS, see \citet[e.g.][]{Miller2019PSRJ0030}.
Recent observational advances have enabled many new studies.
For example, 
the first detection of a binary neutron-star merger by LIGO/VIRGO \citep{Abbott2017GW170817} combined with the associated short gamma-ray burst (sGRB), 
kilonova, and afterglow 
led to an inferred upper limit of the allowed NS mass, $M_{\mathrm{max}}$, of 2.3 -- 2.4~\msun 
\citep{MargalitMetzger2017MaxNS,Ruiz2018MaxNSMass,Rezzolla2018MaxNSMass,Shibata2019MaxNSMass}. 
Bayesian parameter estimation on the combined observations of GW170817, AT2017gfo, and GRB170817A provides a multi-messenger constraint on the tidal deformability, the total mass, and mass-radius pair for GW170817 \citep{Coughlin2019Bayes}.
Light curve modeling of X-ray burst GS~1826-24 constrains the stiffness of the nuclear EOS through the monotonic relationship between the NS radius and the peak luminosity of the burst \citep{Dohi2021XRB}.

Although there are large uncertainties \citep{Even2020Composition,Korobkin2020Axisymmetric,Zhu2020KilonovaUncertainties}, modeling of the electromagnetic radiation of the sGRB and of the kilonova constitutes yet another complimentary constraint on the EOS \citep[e.g.][]{Nedora2021GW170817Model}.

The nuclear EOS is a crucial ingredient in numerical simulations of compact binary mergers and core-collapse supernovae. Since the precise physics is unknown, modelers parameterize over their ignorance and provide tabulated EOSs, which are used as an input parameter \citep{BurgioEOSReview}.

In this work, we focus on CCSNe, where the EOS can have a significant impact on both the dynamics and the outcome. When a massive star runs out of nuclear fuel, it collapses under its own weight. The inner part of the iron core of the star becomes ultra dense and neutron rich, forming a \textit{proto-neutron star} (PNS). In-falling material then bounces off of the PNS, forming an outward moving shock. If this shock is strong enough, it moves outward, driving an explosion.
If the shock fails, the star collapses without an explosion. The core of the star eventually forms either a stable NS, supported by neutron degeneracy pressure and the strong nuclear force, or collapses to a black hole (BH) \cite{shapiro2008black}.

Solving the full CCSN problem remains a grand challenge in astrophysics, with many decades devoted to simulations at high resolution in 3D \citep[][e.g.]{Muller2020LRCA}. Several studies have been performed with a focus on the EOS dependence of the PNS contraction and explosion properties, \citet[e.g.][]{Richers2017EOS,Yasin2020EOS,Schneider2019EOS}.
Many other studies performed simulations of 1-3 different progenitors with 1-3 different EOSs. See \citet{Nakazato2021EOS,IvanovFernandez2021EOS} for some recent examples. 
Similar small-scale sensitivity studies have also been attempted for binary merger simulations.
However, the absence of a truly systematic approach makes it difficult to correlate the explosion properties with the underlying physics assumptions in the EOS.

%% THIS PAPER - OUR WORK %%%
In this letter, we take an entirely new approach, facilitated by the availability of computationally relatively inexpensive and effective CCSN models. 
We combine the CCSN simulations with statistical techniques from data science and observational data of NS and BH remnant masses and use a data-driven framework to validate six nuclear EOS models. 

Unfortunately, with rare exceptions \citep{vanDyk.SNprogenitors}, it is impossible to connect a supernova explosion to its progenitor star.
Rather, what \emph{is} available is a set of observations of living stars (not yet collapsing), a separate set of observations of explosions, and another set of observations of post-explosion remnant objects (BHs or NSs).
This enables us to estimate the \emph{probability distributions} of these populations, but without any means to draw a direct connection between a specific progenitor, a specific supernova, and a specific remnant.

We sidestep this issue by comparing distributions directly.
We use a suite of CCSN models to map a physically motivated statistical distribution of stellar progenitors into a synthetic distribution of remnants, which can then be compared to observed remnant distributions.
We create \emph{six} synthetic distributions, one for each EOS, and use this comparison to validate the EOS models.

%%%%%%%%%%%%%%%%%%%%%%%%%%%%%%%%%%%%%%%%%%%%%%%%
\section{Methodology and Input Physics} \label{sec:method}
%%%%%%%%%%%%%%%%%%%%%%%%%%%%%%%%%%%%%%%%%%%%%%%%

\begin{figure*}
\centering
    \includegraphics[scale=0.35]{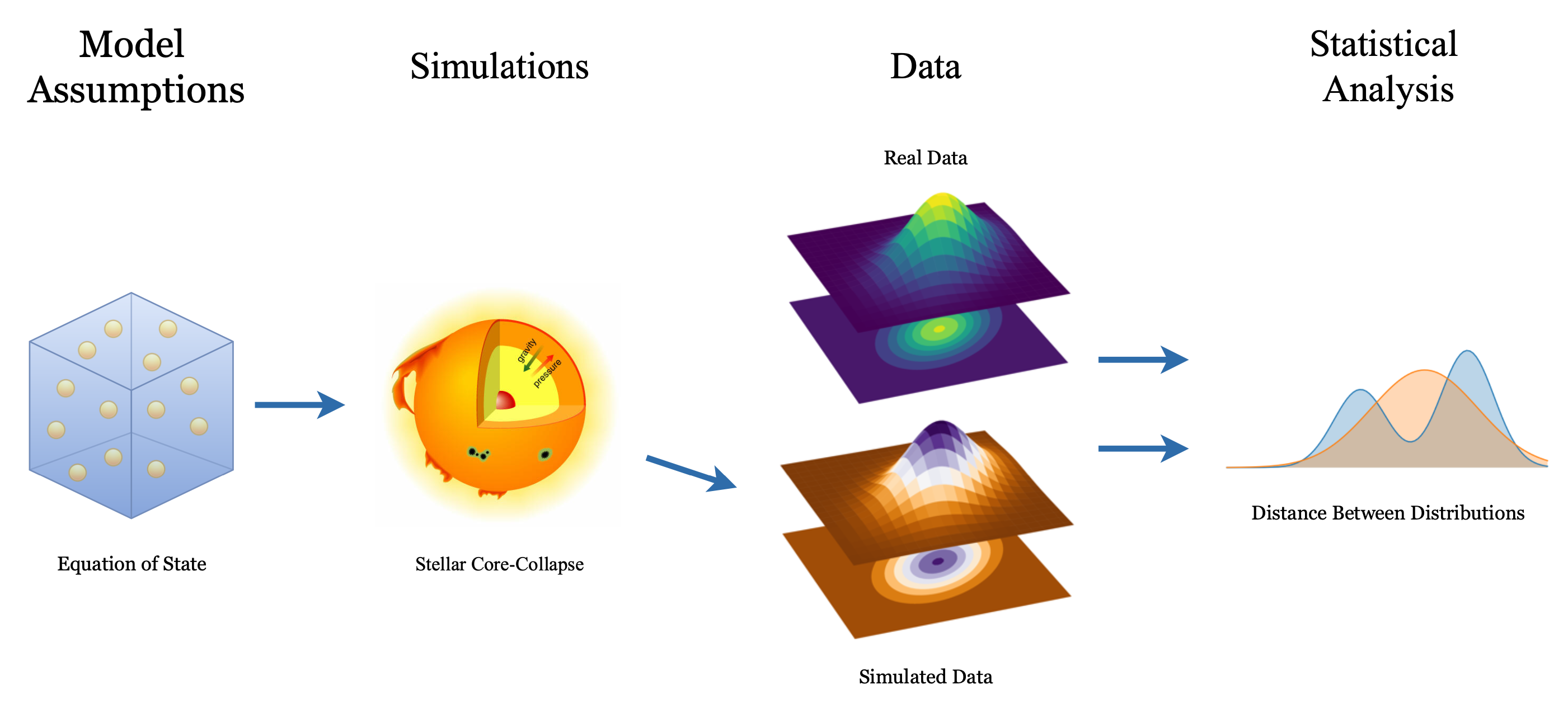}
    \caption{An overview of our methodology. Different equations of state lead to differences in simulated outcomes of core-collapse. 
    \label{fg:overview}
    }
\end{figure*}

An overview of our general methodology is shown in Figure~\ref{fg:overview}. 
We begin with a set of model assumptions embedded in different EOSs that lead to different simulations of stellar collapse. These are used to generate a statistical distribution of post-collapse remnant objects. 
The simulated distribution is then compared to an observed distribution of remnants. A quantification of the distance between real and simulated distributions points to a favored model assumption.

% ----- Figure: Project Overview -----------------------------
% -----------------------------------------------------

The distribution of stellar progenitors depends on the zero age main sequence mass $\mzams$ and metalicity $z$, which we combine into the single input variable $\bx$. 
Our simulations vary over $\bx$ and EOS $s$. A given simulation predicts as output either a NS or BH of a given mass. We call this simulated mapping from $\bx$ to a remnant mass $\eta(\bx; s)$, $\zeta(\bx)$ --which is assumed to mimic the true mapping found in Nature. 
The observed data is also subject to measurement error, $e$; this error emerges from limitations of the observed system or the observing instrument.

%%%%%%%%%%%%%%%%%%%%%%%%%%%%%%%%%%%%%%%%%%%%%%%%
\subsection{Distribution of Mass and Metallicity}
\label{subsec:IMF}

For pre-collapse stars, we use the initial mass function from \citet{Kroupa}:
\begin{small}
\begin{equation}
    \label{eq:IMF}
    \xi(\mzams, z) = \begin{cases}
    0.035 \mzams^{-1.3}&\text{for }\mzams < 0.5\\
    0.019 \mzams^{-2.2}&\text{for }0.5 \leq \mzams < 1.0\\
    0.019 \mzams^{-2.7}&\text{for } \mzams \geq 1.0
    \end{cases},
\end{equation}
\end{small}
for a ZAMS mass $\mzams$ in solar masses (\msun) and metallicity $z$ in units of solar metallicity. Note that the distribution is uniform in metallicity and that it 
may need a global renormalization factor.

%%%%%%%%%%%%%%%%%%%%%%%%%%%%%%%%%%%%%%%%%%%%%%%%
\subsection{Simulations} \label{subsec:simulated_data}

While computational power and model sophistication are rapidly improving, three-dimensional CCSN simulations are still heroic, computationally-expensive endeavors,
making a large-scale parameter study intractable. 
Here, we make use of a more tractable setup, where the explosion is driven in spherical symmetry by a self-consistent parameterized treatment \citep{PUSH1}. 
Our framework (``PUSH'') mimics the net enhanced energy deposition expected from multi-dimensional fluid motion in a spherically symmetric simulation. 

We have performed 681 simulations with initial masses of $\mzams = 10.8$ -- $40$~\msun at three initial metallicities\footnote{A larger initial mass range is desirable, but the relevant progenitor models are currently unavailable to us. As a consequence, we may miss some of the lowest mass neutron stars in our analysis.} ($z=0, 10^{-4},1$~$Z_{\odot}$) \citep{02models,07models} and six different nuclear EOS models (`states'). The outcome ('output') is either a NS (in 508 cases) or a BH (in 173 cases). The sample size per EOS is $\sim 100$ ($\sim 180$ for DD2).

% EOS description (physics)
The EOS models are taken from \citet{Steiner.Hempel.Fischer.eos:2013} (SFHo, SFHx), from \citet{hempel2010,hempel2012} (HS(DD2), HS(TM1), HS(NL3)), and from \citet{banik2014} (BHB$\lambda\phi$). All EOS models have $M_{\mathrm{max}} > 2 M_{\odot}$ for a cold NS and are based on the relativistic mean field model with four different parametrizations (DD2, SFH, NL3, TM1). 
The BHB$\lambda\phi$ EOS additionally includes hyperons at high densities, which softens the EOS.

%%%%%%%%%%%%%%%%%%%%%%%%%%%%%%%%%%%%%%%%%%%%%%%%
\subsection{Distributions of Observed Remnant Mass}
\label{subsec:true_data}

Distributions of the populations of isolated NSs, binary NS systems, and BHs are all available from observations. 
Since it is not currently possible to combine these into a joint distribution, we examine simulated BH and simulated NS populations separately, and compare them to observed populations on a population-by-population basis. Furthermore, we compare the simulated NS population to each observed NS population separately, since the simulator cannot distinguish between isolated and binary remnants.

For the observed BH mass distribution, we use model C from \citet{Abbott19}, which was first proposed in \citet{TalbotThrane2018}. Model C predicts a smooth transition from no BHs below some mass $\bzbh^-$ to a truncated power law that vanishes at $\bzbh^+$, and a Gaussian distribution of high-mass BHs formed from pair-instability supernovae. Since our models do not capture the pair instability, we drop the second Gaussian peak of the distribution, resulting in a smooth, truncated power law:
\begin{equation}
    \label{eq:BH:mass}
    p(\bzbh) = C \bzbh^{-\alpha} \Theta(\bzbh^+ - \bzbh) S(\bzbh^-, \delta_m)
\end{equation}
\noindent
where the BH mass $\bzbh$ is in solar masses. Here $C$ is a normalization constant, $\Theta$ is the Heaviside step function, and $S$ is a smoothing function that goes from zero at $\bzbh$ to unity at $\bzbh + \delta_m$.\footnote{The formal definition can be found in \citet{TalbotThrane2018}, but if smoothness is not important, a simple line with slope $\delta_m$ suffices.} We set the other parameters to those found in \citet{Abbott19}: $\alpha=7.1$, $\bzbh^{-}=6.8$, $\bzbh^{+}=75$, $\delta_m = 3$.

For the observed NS masses, we consider four observed populations of neutron stars: slow pulsars (NSS), recycled pulsars (NSR), double-degenerate neutron stars in circular orbits (NSC), and double-degenerate neutron stars in eccentric obits (NSE). Here, single-degenerate refers to a NS in a binary with a non-degenerate companion, double-degenerate to a NS in a binary with another NS.
All four are sampled from a Gaussian \citep{Ozel}
\begin{equation}
    \label{eq:NS:Mass}
    p(\bzns) = \frac{1}{\sqrt{2 \pi \sigma^2}} e^{ -(\bzns - M_0)^2/(2 \sigma^2)},
\end{equation}
for NS mass $M_{\mathrm{NS}}$. 
The parameters are: $M_0 = 1.49$ and $\sigma=0.19$ for NSS, $M_0=1.54$ and $\sigma=0.23$ for NSR, $M_0=1.33$ and $\sigma=0.09$ for NSC, and $M_0=1.29$ and $\sigma=0.24$ for NSE.  All units are in solar masses.

%%%%%%%%%%%%%%%%%%%%%%%%%%%%%%%%%%%%%%%%%%%%%%%%
\subsection{Distributions of Measurement Error}
\label{subsec:error}

The observed mass distributions are subject to observational error coming from, e.g., 
telescope properties and the difficulty of observing a given system. We convert an ensemble of measurements to a distribution by assuming, via the central limit theorem \citep{fischer2010history}, that in the ensemble, the error $e$ can be drawn from a 
Gaussian distribution with mean zero and standard deviation $\sigma$ that depends on the population:
\begin{equation}
    \label{eq:gaussian}
    p(e;\sigma) = \frac{1}{\sqrt{2\pi\sigma^2}} e^{-e^2/2\sigma^2},
\end{equation}

The BH measurement error comes primarily from the noise properties of gravitational wave detectors, which depends on the frequency of the gravitational wave, and thus the masses of the merging BHs \citep{LigoNoise}. Lower mass black holes produce higher-frequency gravitational waves, which spend a longer time in the LIGO band and thus the detector is able to build up a better signal-to-noise ratio, resulting in a lower error.
From reported 90\% confidence intervals \citep{Abbott19}, we infer the standard deviation of the measurement error to be
\begin{equation}
    \label{eq:sigma:bh:m}
    \sigma_{\mathrm{BH}}(\bzbh) = 0.120213 \bzbh + 0.355936
\end{equation}
For NSs, observations vary significantly in their sensitivity depending on observing telescope and observed system. We therefore assume the standard deviation for the neutron star mass observational error (in solar masses) to be
\begin{eqnarray}
    \label{eq:sigma:ns:m:NSS}
    \sigma_{\mathrm{NSS}} &=& 0.12\\
    \label{eq:sigma:ns:m:NSR}
    \sigma_{\mathrm{NSR}} &=& 0.088\\
    \label{eq:sigma:ns:m:NSC}
    \sigma_{\mathrm{NSC}} &=& 0.003\\
    \label{eq:sigma:ns:m:NSE}
    \sigma_{\mathrm{NSE}} &=& 0.28
\end{eqnarray}
We arrive at this value by taking the average width of 90\% confidence intervals of observations for the appropriate populations.

%%%%%%%%%%%%%%%%%%%%%%%%%%%%%%%%%%%%%%%%%%%%%%%%
\section{Statistical Techniques}
\label{sec:statistics}

Although our CCSN models contain many uncertainties, for the purposes of this study, we assume that our simulations would mimic physical reality closely
if $s$ is tuned to an unknown ideal EOS $s_0$:
\begin{equation}
\zeta(\bx) \approx \eta(\bx;s_0), \quad \forall \bx.
\end{equation}
In other words, we assume that the discrepancy between the best tuned simulator $\eta(\cdot;s_0)$ and the reality $\zeta(\cdot)$ is negligible. 
The setup assuming discrepancy between $\eta(\cdot;s_0)$ and $\zeta(\cdot)$ suffers from an identifiability issue \citep[e.g.,][]{Brynjarsdottir-OHagan14,Wong-Storlie-Lee17}, and may require additional information to find $s_0$. In the following, we discuss the predicted and observed remnant mass distributions for BHs, $\bzbh$ and $(\bzbh)^s$ respectively. The same procedure also applies to NSs $\bzns$ and $(\bzns)^s$, respectively. 

Our overarching goal is to find the state $s_0$, which is only possible if we have observations from Nature. 
However, in practice, we can only observe a contaminated version $\bybh$ (or $\byns$) of $\bzbh$ (or $\bzns$).
The relationship between $\by$ and $\bz$ is modeled by
\begin{equation}
\by = \bz + \be = \zeta(\bx) + \be,
\end{equation}
where $\be$ is the measurement error satisfying $\E(\be\mid \bx) = 0$ and $\bx$ encapsulates both $\mzams$ and $z$.

In a typical setup of computer model calibration, one would collect data in terms of input-output pairs, i.e., $\{((\bx)_i,(\by)_i)\}_{i=1}^n$ from Nature to calibrate the simulator, i.e., to estimate $s_0$, by
\begin{equation}
    \underset{s}{\arg\min}\, \frac{1}{n}\sum_{i=1}^{n}\left\{(\by)_{i} - \eta((\bx)_i;s)\right\}^2,
    \label{optim:common}
\end{equation}
assuming the simulator evaluations are affordable. 
However, here, we do not have the luxury to observe the input-output pair $(\bx, \by)$ from Nature. 
Instead, we are only able to obtain the marginal distribution $p(\bx)$ of $\bx$ (Section \ref{subsec:IMF}) and the marginal distribution $p(\by)$ of $\by$ (Section \ref{subsec:true_data}). Therefore, we have to build a calibration strategy that is intuitively based on $p(\by)$ and $p(\bx)$, instead of the joint distribution $p(\bx, \by)$ or the conditional distribution $p(\by\mid \bx)$.

%%%%%%%%%%%%%%%%%%%%%%%%%%%%%%%%%%%%%%%%%%%%%%%%
\subsection{Comparing the marginal distributions}

Aside from real star-collapse data, we are additionally provided with data sets $\{((\bx )^s_{i}, (\bz)^s_{i})\}_{i=1}^{n_s}$ from the simulator for every state $s$:
\begin{equation}
    (\bz)^s_{i} = \eta((\bx)^s_{i}; s), \quad i=1,\dots, n_s,
\end{equation}
where the sampling designs $\{(\bx)^s_{i}\}$ are generated independently according to the sampling density $q_s(\bx)$.

We use $q_s$ to represent the densities related to data generated from the simulator at state $s$: e.g., $q_s(\bz,\bx)$, $q_s(\bx)$ and $q_s(\bz\mid \bx)$. Note that
\begin{align}
    p(\by) &= \int p(\bx,\by) \mathrm{d}\bx\\
    &= \int p(\by \mid \bx) p(\bx) \mathrm{d}\bx\\
    &= \int p(\by \mid \bx)w_s(\bx) q_s(\bx) \mathrm{d}\bx, \label{eqn:py}
\end{align}
where $w_s(\bx) := p(\bx)/q_s(\bx)$ is assumed to be strictly positive and known.

We define the contaminated version of $(\bz)^{s}$ as $(\by)^{s}:=(\bz)^{s} + (\be)^{s}$  where $(\be)^{s} \mid (\bx)^s \sim p(\be \mid \bx)$. Again, we use $q_s$ to represent related densities: e.g., $q_s(\by,\bx)$ and $q_s(\by\mid \bx)$.
 We have
\begin{align}
    p_s(\by) :=& \int q_s(\by,\bx) w_s(\bx) \mathrm{d}x\\
    =& \int q_s(\by\mid \bx) q_s(\bx) w_s(\bx) \mathrm{d}\bx\label{eqn:psy}
\end{align}
Since $q_{s_0}(\by \mid \bx) = p(\by \mid \bx)$, $p_{s_0}(\by) = p(\by)$ by \eqref{eqn:py} and \eqref{eqn:psy}.
We then estimate $p(\by)$ through a weighted kernel density estimation (KDE) of $p_{s_0}(\by)$.
The weighted KDE is based on the Fast Fourier Transform algorithm implemented in KDEpy \citep{tommy_odland_2018_2392268}.
The pseudo code of the estimation of $p_s$ can be found in Appendix \ref{app:ps}.

Assuming $q_s\neq q_{s_0}$ for any $s\neq s_0$, we can compare $p(\by)$ with the weighed KDE at different $s$, and choose $s_0$ as the ``closest'' one.
We use the following two information-theoretic statistical distances to quantify the dissimilarity between $p(\cdot)$ and the estimated $p_s(\cdot)$.
The first is the Kullback-Leibler divergence
\begin{equation}
  D_{\mathrm{KL}}(p\, ||\, p_s) = \int p(\by)\log\left(\frac{p(\by)}{p_s(\by)}\right) \mathrm{d}\by,
\end{equation}
that measures the relative entropy between $p(\by)$ and $p_s(\by)$.
The second is the total variation distance $D_{\mathrm{TV}}$
\begin{equation}
    D_\mathrm{TV}(p, p_s) = \frac{1}{2}\int|p(\by) - p_s(\by)|\mathrm{d}\by,
\end{equation}
that measures the maximum difference between the probabilities assigned to an event by two probability distributions.

We also report on the corresponding 95\% confidence intervals via bootstrap quantiles \citep{efron1994introduction}.
The pseudo code of the bootstrap procedure is given in Appendix \ref{app:bootstrap}.

%%%%%%%%%%%%%%%%%%%%%%%%%%%%%%%%%%%%%%%%%%%%%%%%
\section{Results} \label{sec:results}
%%%%%%%%%%%%%%%%%%%%%%%%%%%%%%%%%%%%%%%%%%%%%%%%

\begin{figure}[t]
    \centering
    \includegraphics[width=0.49\textwidth]{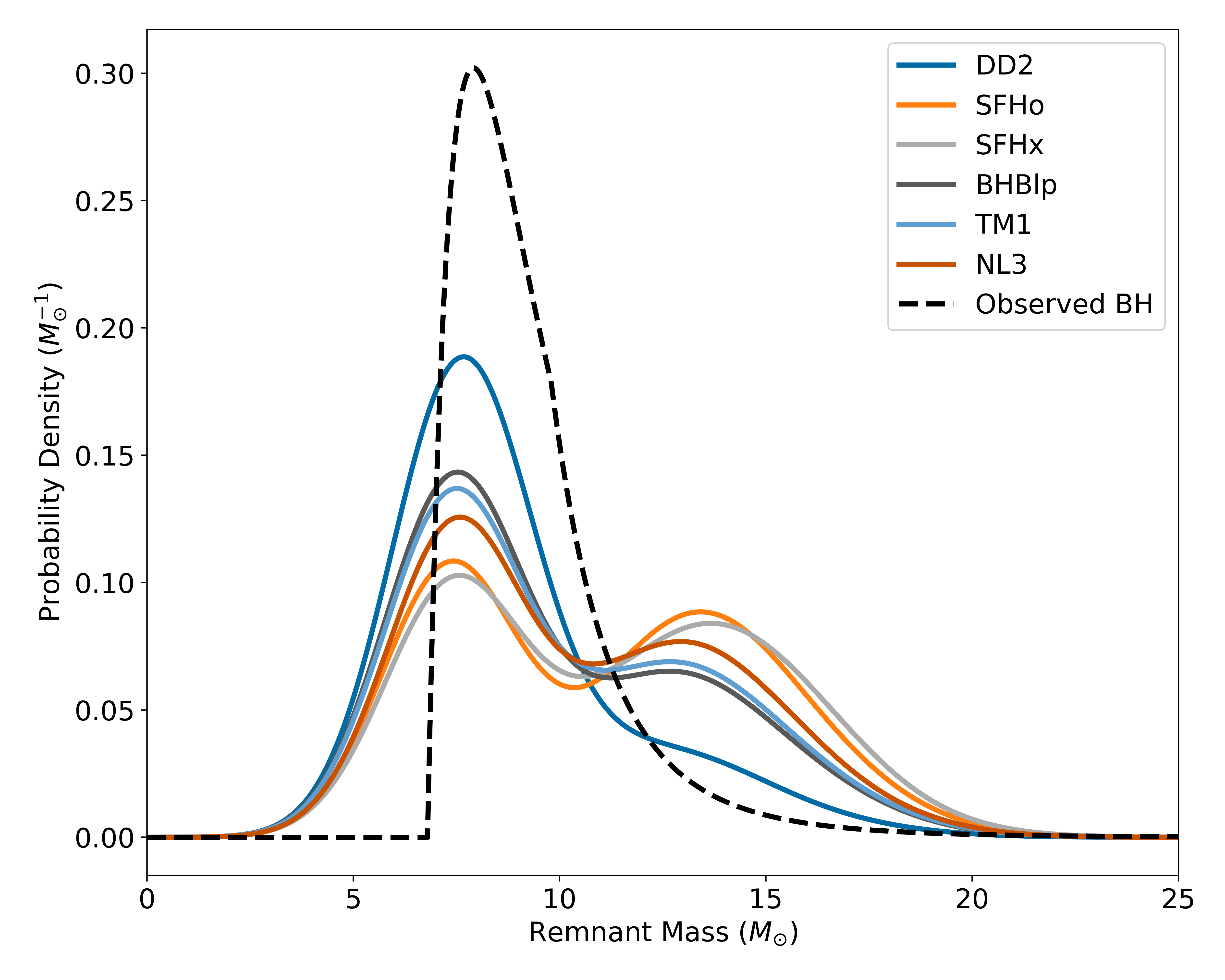}
    \includegraphics[width=0.49\textwidth]{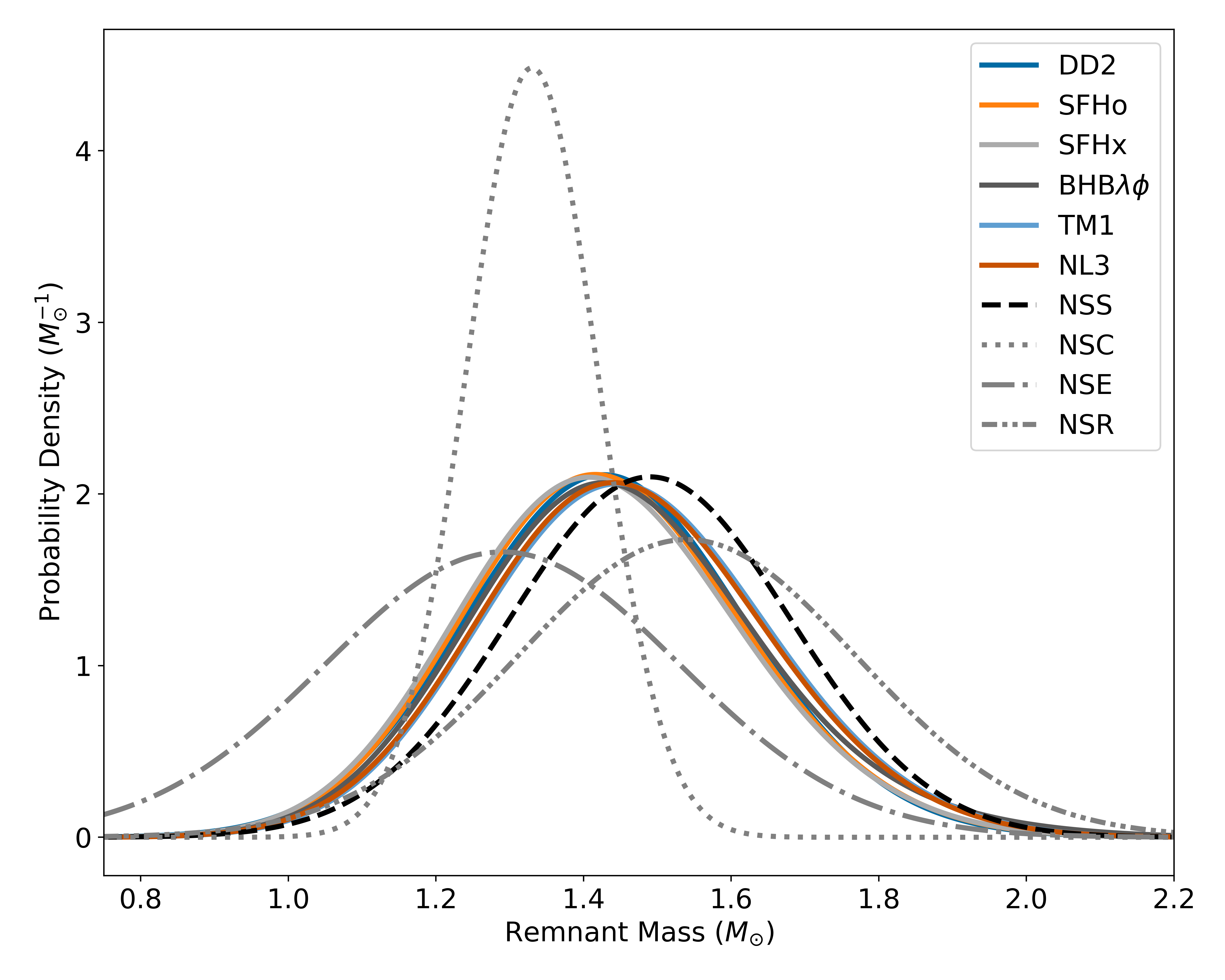}
    \caption{Probability density functions of BHs (top) and NS (bottom) for simulated (color) and observed(black; dashed for isolated and dotted for double NS) distributions. 
    \label{fig:pdfs}
    }
\end{figure}

The simulated  and observed distributions of remnant masses (BHs and NSs separately) are shown in Figure \ref{fig:pdfs}.
We find the simulated neutron star population best agrees with the observed NSS population. We believe this is reasonable for the following reasons: The PUSH models use isolated stars as initial conditions, and double-degenerate binaries likely deviate more from single-star evolution than do single-degenerate. Also, recycled pulsars have accreted enough mass and angular momentum to spin up, meaning their mass reflects more than supernova dynamics. We therefore do not compare distances for the NSR, NSC, or NSE populations.

\begin{deluxetable}{lllllll}
\tablecaption{
    Summary of distances for all three populations and all six EOS models.
    \label{tab:distances}
}
\tablewidth{0pt}
\tablehead{
    \colhead{EOS} &
    \twocolhead{BH} &
    \twocolhead{NSS} \\
    \colhead{} & 
    \colhead{$D_{KL}$} & 
    \colhead{$D_{TV}$} &
    \colhead{$D_{KL}$} & 
    \colhead{$D_{TV}$}
}
\startdata
DD2 &
    0.528$^{+0.101}_{-0.068}$ &  % KL for BH
    0.324$^{+0.050}_{-0.036}$ &  % TV for BH
    0.066$^{+0.058}_{-0.043}$ &  % KL for NSS
    0.125$^{+0.045}_{-0.052}$ \\ % TV for NSS
SFHo &
    1.011$^{+0.836}_{-0.270}$ &  % KL for BH
    0.523$^{+0.155}_{-0.092}$ &  % TV for BH
    0.071$^{+0.081}_{-0.054}$ &  % KL for NSS
    0.133$^{+0.060}_{-0.069}$ \\ % TV for NSS
SFHx &
    0.999$^{+1.510}_{-0.312}$ &  % KL for BH
    0.519$^{+0.187}_{-0.109}$ &  % TV for BH
    0.082$^{+0.098}_{-0.063}$ &  % KL for NSS
    0.143$^{+0.066}_{-0.075}$ \\ % TV for NSS
BHB$\lambda\phi$ &
    0.729$^{+0.270}_{-0.135}$ &  % KL for BH
    0.427$^{+0.089}_{-0.060}$ &  % TV for BH
    0.044$^{+0.066}_{-0.033}$ &  % KL for NSS
    0.105$^{+0.061}_{-0.064}$ \\ % TV for NSS
TM1 &
    0.752$^{+0.295}_{-0.152}$ &  % KL for BH
    0.437$^{+0.093}_{-0.065}$ &  % TV for BH
    0.018$^{+0.056}_{-0.016}$ &  % KL for NSS
    0.068$^{+0.068}_{-0.050}$ \\ % TV for NSS
NL3 &
    0.808$^{+0.459}_{-0.178}$ &  % KL for BH
    0.457$^{+0.121}_{-0.073}$ &  % TV for BH
    0.025$^{+0.060}_{-0.023}$ &  % KL for NSS
    0.080$^{+0.066}_{-0.062}$ \\ % TV for NSS
\enddata
%\tablecomments{
%    Note that we exclude from our analysis the BH formed with SFHx due to the small number of simulations with this EoS and outcome.}
\end{deluxetable}

The distances from each simulated population to the observed one for each distance metric are summarized in Table~\ref{tab:distances}.
We note that for BHs, the $D_{TV}$ distances are similar in magnitude to those for NSSs, whereas the $D_{KL}$ distances for BHs are smaller than those for NSSs. Within each population, the $D_{KL}$ and the $D_{TV}$ give the same ranking, however the rankings are slightly different between NSs and BHs.

\begin{figure*}[htbp]
    \centering
    \includegraphics[width=\textwidth]{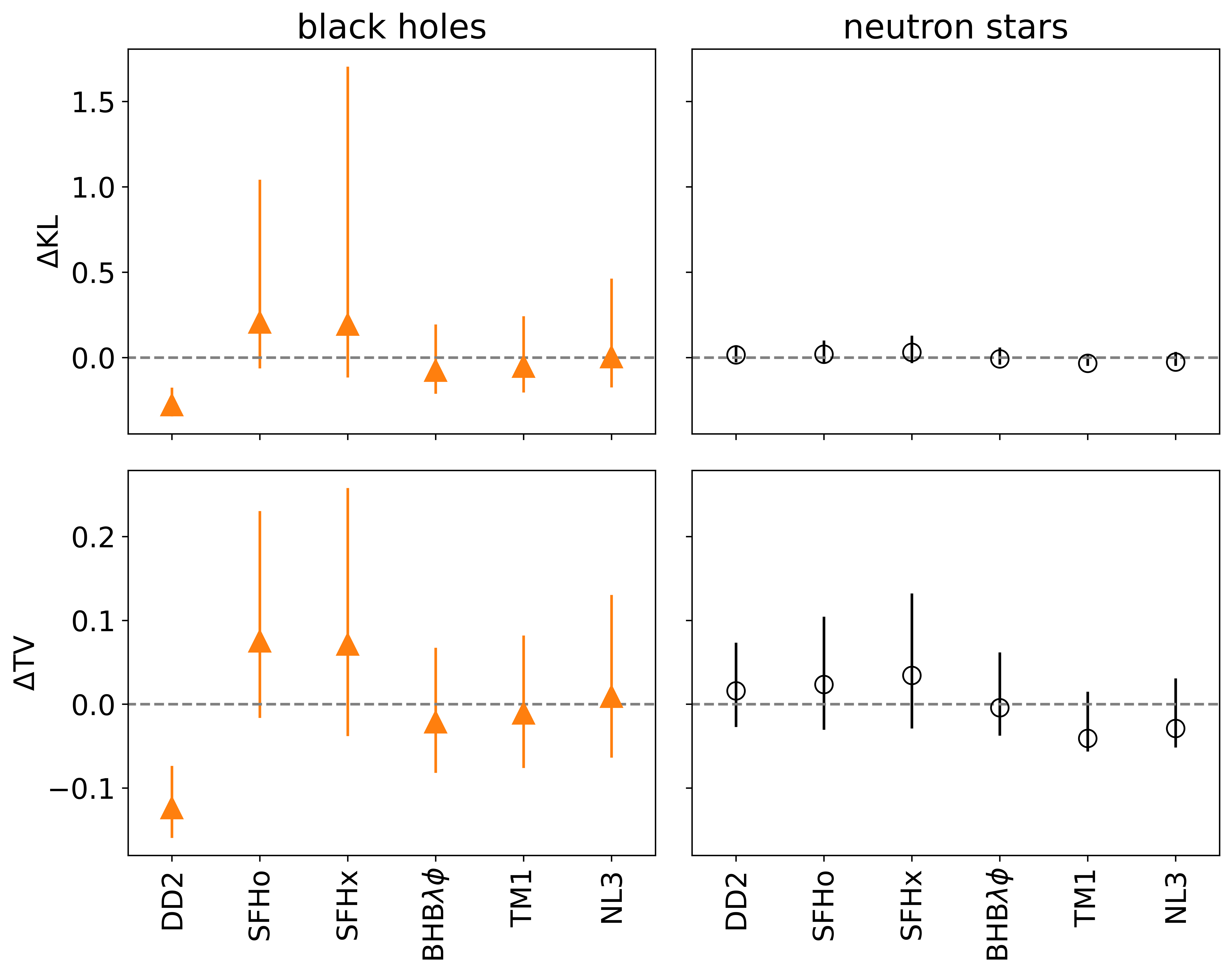}
    \caption{The distance from the mean for each EOS and each population for $D_{KL}$ (top) and $D_{TV}$ (bottom) for BHs (left) and isolated NSs (right). 
    Error bars show 95\% confidence intervals.
    \label{fig:delta:distances}
    }
\end{figure*}

To interpret the results further, we compute for each population, for each distance measure ($D_{KL}$, $D_{TV}$), and for each EOS, the difference from the average distance, $\Delta \text{KL}$ and $\Delta \text{TV}$, as shown in Figure \ref{fig:delta:distances}. The error bars represent 95\% confidence intervals. Points below the dashed line are more favored than points above. 
The BHs and the NSs agree qualitatively except for DD2, which more favored by BHs and less favored by NSs. SHFo and SFHx are both mildly disfavored by all metrics, while $BHB\lambda\phi$, TM1, and NL3  appear to be mildly more favored. 

We remark that the KL divergence sometimes suffers from stability issues due to small density values, such as the tails of our distributions.

%%%%%%%%%%%%%%%%%%%%%%%%%%%%%%%%%%%%%%%%%%%%%%%%
\section{Conclusions} \label{sec:conclusions}
%%%%%%%%%%%%%%%%%%%%%%%%%%%%%%%%%%%%%%%%%%%%%%%%

Future work will explore how machine learning can play the role of a simulator surrogate that provides additional data to improve our statistical confidence during the assessment of theoretical equations \cite{Kamdar15,Behler16}. The learning model can then be combined with active learning \cite{Settles12} to choose examples (feature vectors) in the input space that maximize model performance.  The goal is to avoid the computational cost attached to many simulations. Additionally, physical constraints can be incorporated into the learning model to further narrow the parameter space.

On the astrophysics side, our PUSH models assume stellar progenitors for isolated stars, when in fact most stars are born in binaries, and the observed BH and NS populations reflect this reality. A straightforward generalization would be to include progenitors from binaries---at least at the state of the art of stellar evolution \citep{bsg_models}. Indeed, the choice of weights for stellar progenitors is relatively simple. A more sophisticated approach uses binary population synthesis forward modeling \cite{IzzardBook} to predict the pdf for binary systems. Additionally, 
further work is required to understand the influence of the calibration procedure on our results.
The PUSH method was calibrated using the DD2 EOS \citep{PUSH1}. However, as discussed in \citet{PUSH2}, the PUSH parameters are partially constrained by observations, independent of equation of state, giving some confidence that the EOS used for calibration is subdominant in the procedure.
We remark that the NS masses obtained with PUSH are very similar to the NS masses obtained from 3D simulations in \citet{Burrows2020}, giving confidence that the results of this work qualitatively extend to other supernova models.
This work also used the observed distributions of remnant mass only; this neglects important information about the EOS from the NS radius. However, additional distributions, such as explosion energy and nickel mass, are available in both observations and simulations, providing an exciting direction of multivariate distributions with additional independent variables.
Finally, a more sophisticated construction of the measured populations and associated measurement errors would be a natural extension of our work.

\begin{acknowledgements}
We would like to thank Matthias Hempel for the EOS tables and for fruitful discussions. JMM would like to thank Chris Fryer, Josh Dolence, Ben Ryan, Greg Salvesen, Carl Fields, and Soumi De for useful feedback. We also express our gratitude to the anonymous referee for many constructive comments, improving the quality of the manuscript.
The work at NC State was supported by United States Department of Energy (DOE), Office of Science, Office of  Nuclear Physics under Award DE-FG02-02ER41216. 
Research presented in this work was supported by the Laboratory Directed Research and Development program of Los Alamos National Laboratory under project numbers 20190021DR and 20220087DR. Los Alamos National Laboratory is operated by Triad National Security, LLC, for the National Nuclear Security Administration of U.S. Department of Energy (Contract No. 89233218CNA000001). This work is approved for unlimited release, with number LA-UR-21-30803.
The work at Texas A\&M University was supported by the National Science Foundations (DMS-1711952 and CCF-1934904).
We are grateful for the support of the Research Computing Data Core at the University of Houston for assistance with the calculations carried out in this work.
\end{acknowledgements}

\software{
Agile \citep{Liebendoerfer.Agile}, 
Python \citep{rossumPythonWhitePaper},
Pandas \citep{reback2020pandas},
NumPy, SciPy %\citep{numpy,scipyLib}, 
Matplotlib %\citep{matplotlib},
and
KDEpy\citep{tommy_odland_2018_2392268}.
}

\bibliography{references}{}

\begin{thebibliography}{}
\expandafter\ifx\csname natexlab\endcsname\relax\def\natexlab#1{#1}\fi
\providecommand{\url}[1]{\href{#1}{#1}}
\providecommand{\dodoi}[1]{doi:~\href{http://doi.org/#1}{\nolinkurl{#1}}}
\providecommand{\doeprint}[1]{\href{http://ascl.net/#1}{\nolinkurl{http://ascl.net/#1}}}
\providecommand{\doarXiv}[1]{\href{https://arxiv.org/abs/#1}{\nolinkurl{https://arxiv.org/abs/#1}}}

\bibitem[{{Abbott} {et~al.}(2017)}]{Abbott2017GW170817}
{Abbott}, B.~P., {et~al.} 2017, \prl, 119, 161101,
  \dodoi{10.1103/PhysRevLett.119.161101}

\bibitem[{Abbott {et~al.}(2019)}]{Abbott19}
Abbott, B.~P., {et~al.} 2019, The Astrophysical Journal, 882, L24

\bibitem[{{Banik} {et~al.}(2014){Banik}, {Hempel}, \&
  {Bandyopadhyay}}]{banik2014}
{Banik}, S., {Hempel}, M., \& {Bandyopadhyay}, D. 2014, \apjs, 214, 22,
  \dodoi{10.1088/0067-0049/214/2/22}

\bibitem[{Behler(2016)}]{Behler16}
Behler, J. 2016, J. Chem. Phys., 145, 170901

\bibitem[{Brynjarsd{\'o}ttir \& O'Hagan(2014)}]{Brynjarsdottir-OHagan14}
Brynjarsd{\'o}ttir, J., \& O'Hagan, A. 2014, Inverse problems, 30, 114007

\bibitem[{Buikema {et~al.}(2020)Buikema, Cahillane, Mansell, Blair, Abbott,
  Adams, Adhikari, Ananyeva, Appert, Arai, Areeda, Asali, Aston, Austin, Baer,
  Ball, Ballmer, Banagiri, Barker, Barsotti, Bartlett, Berger, Betzwieser,
  Bhattacharjee, Billingsley, Biscans, Blair, Bode, Booker, Bork, Bramley,
  Brooks, Brown, Cannon, Chen, Ciobanu, Clara, Cooper, Corley, Countryman,
  Covas, Coyne, Datrier, Davis, Di~Fronzo, Dooley, Driggers, Dupej, Dwyer,
  Effler, Etzel, Evans, Evans, Feicht, Fernandez-Galiana, Fritschel, Frolov,
  Fulda, Fyffe, Giaime, Giardina, Godwin, Goetz, Gras, Gray, Gray, Green,
  Gustafson, Gustafson, Hanks, Hanson, Hardwick, Hasskew, Heintze,
  Helmling-Cornell, Holland, Jones, Kandhasamy, Karki, Kasprzack, Kawabe,
  Kijbunchoo, King, Kissel, Kumar, Landry, Lane, Lantz, Laxen, Lecoeuche,
  Leviton, Liu, Lormand, Lundgren, Macas, MacInnis, Macleod, M\'arka, M\'arka,
  Martynov, Mason, Massinger, Matichard, Mavalvala, McCarthy, McClelland,
  McCormick, McCuller, McIver, McRae, Mendell, Merfeld, Merilh, Meylahn,
  Mistry, Mittleman, Moreno, Mow-Lowry, Mozzon, Mullavey, Nelson, Nguyen,
  Nuttall, Oberling, Oram, O'Reilly, Osthelder, Ottaway, Overmier, Palamos,
  Parker, Payne, Pele, Penhorwood, Perez, Pirello, Radkins, Ramirez,
  Richardson, Riles, Robertson, Rollins, Romel, Romie, Ross, Ryan, Sadecki,
  Sanchez, Sanchez, Saravanan, Savage, Schaetzl, Schnabel, Schofield, Schwartz,
  Sellers, Shaffer, Sigg, Slagmolen, Smith, Soni, Sorazu, Spencer, Strain, Sun,
  Szczepa\ifmmode~\acute{n}\else \'{n}\fi{}czyk, Thomas, Thomas, Thorne,
  Toland, Torrie, Traylor, Tse, Urban, Vajente, Valdes, Vander-Hyde, Veitch,
  Venkateswara, Venugopalan, Viets, Vo, Vorvick, Wade, Ward, Warner, Weaver,
  Weiss, Whittle, Willke, Wipf, Xiao, Yamamoto, Yu, Yu, Zhang, Zucker, \&
  Zweizig}]{LigoNoise}
Buikema, A., Cahillane, C., Mansell, G.~L., {et~al.} 2020, Phys. Rev. D, 102,
  062003, \dodoi{10.1103/PhysRevD.102.062003}

\bibitem[{{Burrows} {et~al.}(2020){Burrows}, {Radice}, {Vartanyan}, {Nagakura},
  {Skinner}, \& {Dolence}}]{Burrows2020}
{Burrows}, A., {Radice}, D., {Vartanyan}, D., {et~al.} 2020, \mnras, 491, 2715,
  \dodoi{10.1093/mnras/stz3223}

\bibitem[{{Coughlin} {et~al.}(2019){Coughlin}, {Dietrich}, {Margalit}, \&
  {Metzger}}]{Coughlin2019Bayes}
{Coughlin}, M.~W., {Dietrich}, T., {Margalit}, B., \& {Metzger}, B.~D. 2019,
  \mnras, 489, L91, \dodoi{10.1093/mnrasl/slz133}

\bibitem[{{Curtis} {et~al.}(2020){Curtis}, {Wolfe}, {Fr{\"o}hlich}, {Miller},
  {Wollaeger}, \& {Ebinger}}]{CurtisLightCurves}
{Curtis}, S., {Wolfe}, N., {Fr{\"o}hlich}, C., {et~al.} 2020, arXiv e-prints,
  arXiv:2008.05498.
\newblock \doarXiv{2008.05498}

\bibitem[{{Dohi} {et~al.}(2021){Dohi}, {Nishimura}, {Hashimoto}, {Matsuo},
  {Noda}, \& {Nagataki}}]{Dohi2021XRB}
{Dohi}, A., {Nishimura}, N., {Hashimoto}, M., {et~al.} 2021, arXiv e-prints,
  arXiv:2105.13958.
\newblock \doarXiv{2105.13958}

\bibitem[{Ebinger {et~al.}(2020)Ebinger, Curtis, Ghosh, Fröhlich, Hempel,
  Perego, Liebendörfer, \& Thielemann}]{PUSH4}
Ebinger, K., Curtis, S., Ghosh, S., {et~al.} 2020, The Astrophysical Journal,
  888, 91, \dodoi{10.3847/1538-4357/ab5dcb}

\bibitem[{Efron \& Tibshirani(1994)}]{efron1994introduction}
Efron, B., \& Tibshirani, R.~J. 1994, An introduction to the bootstrap (CRC
  press)

\bibitem[{{Even} {et~al.}(2020){Even}, {Korobkin}, {Fryer}, {Fontes},
  {Wollaeger}, {Hungerford}, {Lippuner}, {Miller}, {Mumpower}, \&
  {Misch}}]{Even2020Composition}
{Even}, W., {Korobkin}, O., {Fryer}, C.~L., {et~al.} 2020, \apj, 899, 24,
  \dodoi{10.3847/1538-4357/ab70b9}

\bibitem[{{Fiorella Burgio} \& {Fantina}(2018)}]{BurgioEOSReview}
{Fiorella Burgio}, G., \& {Fantina}, A.~F. 2018, {Nuclear Equation of State for
  Compact Stars and Supernovae}, ed. L.~{Rezzolla}, P.~{Pizzochero}, D.~I.
  {Jones}, N.~{Rea}, \& I.~{Vida{\~n}a}, Vol. 457, 255

\bibitem[{Fischer(2010)}]{fischer2010history}
Fischer, H. 2010, A History of the Central Limit Theorem: From Classical to
  Modern Probability Theory, Sources and Studies in the History of Mathematics
  and Physical Sciences (Springer New York)

\bibitem[{{Hempel} {et~al.}(2012){Hempel}, {Fischer}, {Schaffner-Bielich}, \&
  {Liebend{\"o}rfer}}]{hempel2012}
{Hempel}, M., {Fischer}, T., {Schaffner-Bielich}, J., \& {Liebend{\"o}rfer}, M.
  2012, \apj, 748, 70, \dodoi{10.1088/0004-637X/748/1/70}

\bibitem[{{Hempel} \& {Schaffner-Bielich}(2010)}]{hempel2010}
{Hempel}, M., \& {Schaffner-Bielich}, J. 2010, Nuclear Physics A, 837, 210,
  \dodoi{10.1016/j.nuclphysa.2010.02.010}

\bibitem[{{Ivanov} \& {Fern{\'a}ndez}(2021)}]{IvanovFernandez2021EOS}
{Ivanov}, M., \& {Fern{\'a}ndez}, R. 2021, \apj, 911, 6,
  \dodoi{10.3847/1538-4357/abe59e}

\bibitem[{{Izzard} \& {Halabi}(2018)}]{IzzardBook}
{Izzard}, R.~G., \& {Halabi}, G.~M. 2018, arXiv e-prints, arXiv:1808.06883.
\newblock \doarXiv{1808.06883}

\bibitem[{Kamdar {et~al.}(2015)Kamdar, Turk, \& Brunner}]{Kamdar15}
Kamdar, H.~M., Turk, M.~J., \& Brunner, R.~J. 2015, Monthly Notices of the
  Royal Astronomical Society, 455, 642–658

\bibitem[{{Korobkin} {et~al.}(2021){Korobkin}, {Wollaeger}, {Fryer},
  {Hungerford}, {Rosswog}, {Fontes}, {Mumpower}, {Chase}, {Even}, {Miller},
  {Misch}, \& {Lippuner}}]{Korobkin2020Axisymmetric}
{Korobkin}, O., {Wollaeger}, R.~T., {Fryer}, C.~L., {et~al.} 2021, \apj, 910,
  116, \dodoi{10.3847/1538-4357/abe1b5}

\bibitem[{Kroupa {et~al.}(1993)Kroupa, Tout, \& Gilmore}]{Kroupa}
Kroupa, P., Tout, C.~A., \& Gilmore, G. 1993, Monthly Notices of the Royal
  Astronomical Society, 262, 545

\bibitem[{{Liebend{\"o}rfer} {et~al.}(2001){Liebend{\"o}rfer}, {Mezzacappa}, \&
  {Thielemann}}]{Liebendoerfer.Agile}
{Liebend{\"o}rfer}, M., {Mezzacappa}, A., \& {Thielemann}, F.-K. 2001, \prd,
  63, 104003, \dodoi{10.1103/PhysRevD.63.104003}

\bibitem[{Lonardoni {et~al.}(2020)Lonardoni, Tews, Gandolfi, \&
  Carlson}]{Lonardoni2020}
Lonardoni, D., Tews, I., Gandolfi, S., \& Carlson, J. 2020, Phys. Rev.
  Research, 2, 022033, \dodoi{10.1103/PhysRevResearch.2.022033}

\bibitem[{{Margalit} \& {Metzger}(2017)}]{MargalitMetzger2017MaxNS}
{Margalit}, B., \& {Metzger}, B.~D. 2017, \apjl, 850, L19,
  \dodoi{10.3847/2041-8213/aa991c}

\bibitem[{{Menon} \& {Heger}(2017)}]{bsg_models}
{Menon}, A., \& {Heger}, A. 2017, \mnras, 469, 4649.
\newblock \doarXiv{1703.04918}

\bibitem[{{Miller} {et~al.}(2019){Miller}, {Lamb}, {Dittmann}, {Bogdanov},
  {Arzoumanian}, {Gendreau}, {Guillot}, {Harding}, {Ho}, {Lattimer}, {Ludlam},
  {Mahmoodifar}, {Morsink}, {Ray}, {Strohmayer}, {Wood}, {Enoto}, {Foster},
  {Okajima}, {Prigozhin}, \& {Soong}}]{Miller2019PSRJ0030}
{Miller}, M.~C., {Lamb}, F.~K., {Dittmann}, A.~J., {et~al.} 2019, \apjl, 887,
  L24, \dodoi{10.3847/2041-8213/ab50c5}

\bibitem[{{M{\"u}ller}(2020)}]{Muller2020LRCA}
{M{\"u}ller}, B. 2020, Living Reviews in Computational Astrophysics, 6, 3.
\newblock \doarXiv{2006.05083}

\bibitem[{{Nakazato} {et~al.}(2021){Nakazato}, {Sumiyoshi}, \&
  {Togashi}}]{Nakazato2021EOS}
{Nakazato}, K., {Sumiyoshi}, K., \& {Togashi}, H. 2021, \pasj,
  \dodoi{10.1093/pasj/psab026}

\bibitem[{{Nedora} {et~al.}(2021){Nedora}, {Bernuzzi}, {Radice}, {Daszuta},
  {Endrizzi}, {Perego}, {Prakash}, {Safarzadeh}, {Schianchi}, \&
  {Logoteta}}]{Nedora2021GW170817Model}
{Nedora}, V., {Bernuzzi}, S., {Radice}, D., {et~al.} 2021, \apj, 906, 98,
  \dodoi{10.3847/1538-4357/abc9be}

\bibitem[{Odland(2018)}]{tommy_odland_2018_2392268}
Odland, T. 2018, tommyod/KDEpy: Kernel Density Estimation in Python, v0.9.10,
  Zenodo, \dodoi{10.5281/zenodo.2392268}

\bibitem[{pandas~development team(2020)}]{reback2020pandas}
pandas~development team, T. 2020, pandas-dev/pandas: Pandas, latest,  Zenodo,
  \dodoi{10.5281/zenodo.3509134}

\bibitem[{Perego {et~al.}(2015)Perego, Hempel, Fröhlich, Ebinger, Eichler,
  Casanova, Liebendörfer, \& Thielemann}]{PUSH1}
Perego, A., Hempel, M., Fröhlich, C., {et~al.} 2015, The Astrophysical
  Journal, 806, 275, \dodoi{10.1088/0004-637X/806/2/275}

\bibitem[{{Rezzolla} {et~al.}(2018){Rezzolla}, {Most}, \&
  {Weih}}]{Rezzolla2018MaxNSMass}
{Rezzolla}, L., {Most}, E.~R., \& {Weih}, L.~R. 2018, \apjl, 852, L25,
  \dodoi{10.3847/2041-8213/aaa401}

\bibitem[{{Richers} {et~al.}(2017){Richers}, {Ott}, {Abdikamalov}, {O'Connor},
  \& {Sullivan}}]{Richers2017EOS}
{Richers}, S., {Ott}, C.~D., {Abdikamalov}, E., {O'Connor}, E., \& {Sullivan},
  C. 2017, \prd, 95, 063019, \dodoi{10.1103/PhysRevD.95.063019}

\bibitem[{Rossum(1995)}]{rossumPythonWhitePaper}
Rossum, G. 1995, {P}ython Reference Manual, Tech. rep., Amsterdam, The
  Netherlands, The Netherlands

\bibitem[{{Ruiz} {et~al.}(2018){Ruiz}, {Shapiro}, \&
  {Tsokaros}}]{Ruiz2018MaxNSMass}
{Ruiz}, M., {Shapiro}, S.~L., \& {Tsokaros}, A. 2018, \prd, 97, 021501,
  \dodoi{10.1103/PhysRevD.97.021501}

\bibitem[{{Schneider} {et~al.}(2019){Schneider}, {Roberts}, {Ott}, \&
  {O'Connor}}]{Schneider2019EOS}
{Schneider}, A.~S., {Roberts}, L.~F., {Ott}, C.~D., \& {O'Connor}, E. 2019,
  \prc, 100, 055802, \dodoi{10.1103/PhysRevC.100.055802}

\bibitem[{Settles(2012)}]{Settles12}
Settles, B. 2012, Active Learning (Morgan \& Claypool Publishers)

\bibitem[{Shapiro \& Teukolsky(2008)}]{shapiro2008black}
Shapiro, S., \& Teukolsky, S. 2008, Black Holes, White Dwarfs, and Neutron
  Stars: The Physics of Compact Objects (Wiley)

\bibitem[{{Shibata} {et~al.}(2019){Shibata}, {Zhou}, {Kiuchi}, \&
  {Fujibayashi}}]{Shibata2019MaxNSMass}
{Shibata}, M., {Zhou}, E., {Kiuchi}, K., \& {Fujibayashi}, S. 2019, \prd, 100,
  023015, \dodoi{10.1103/PhysRevD.100.023015}

\bibitem[{{Steiner} {et~al.}(2013){Steiner}, {Hempel}, \&
  {Fischer}}]{Steiner.Hempel.Fischer.eos:2013}
{Steiner}, A.~W., {Hempel}, M., \& {Fischer}, T. 2013, \apj, 774, 17,
  \dodoi{10.1088/0004-637X/774/1/17}

\bibitem[{{Talbot} \& {Thrane}(2018)}]{TalbotThrane2018}
{Talbot}, C., \& {Thrane}, E. 2018, \apj, 856, 173,
  \dodoi{10.3847/1538-4357/aab34c}

\bibitem[{{Van Dyk}(2017)}]{vanDyk.SNprogenitors}
{Van Dyk}, S.~D. 2017, Philosophical Transactions of the Royal Society of
  London Series A, 375, 20160277

\bibitem[{Wong {et~al.}(2017)Wong, Storlie, \& Lee}]{Wong-Storlie-Lee17}
Wong, R. K.~W., Storlie, C.~B., \& Lee, T. C.~M. 2017, Journal of the Royal
  Statistical Society: Series B, 79, 635

\bibitem[{{Yasin} {et~al.}(2020){Yasin}, {Sch{\"a}fer}, {Arcones}, \&
  {Schwenk}}]{Yasin2020EOS}
{Yasin}, H., {Sch{\"a}fer}, S., {Arcones}, A., \& {Schwenk}, A. 2020, \prl,
  124, 092701, \dodoi{10.1103/PhysRevLett.124.092701}

\bibitem[{{Zhu} {et~al.}(2021){Zhu}, {Lund}, {Barnes}, {Sprouse}, {Vassh},
  {McLaughlin}, {Mumpower}, \& {Surman}}]{Zhu2020KilonovaUncertainties}
{Zhu}, Y.~L., {Lund}, K.~A., {Barnes}, J., {et~al.} 2021, \apj, 906, 94,
  \dodoi{10.3847/1538-4357/abc69e}

\bibitem[{Özel \& Freire(2016)}]{Ozel}
Özel, F., \& Freire, P. 2016, Annual Review of Astronomy and Astrophysics, 54,
  401

\end{thebibliography}
\bibliographystyle{aasjournal}

\appendix

 \section{The nuclear equations of state}
 \label{append:eos}

 The following table summarizes the key properties of the nuclear equations of state used in this work.
 \begin{deluxetable}{lllllllll}[h!]
\tablecaption{
    Nuclear matter and neutron star properties of the EoSs used here
    \label{tab:eos} 
}
\tablewidth{0pt}
\tablehead{
    \colhead{`State'} &
    \colhead{EOS} & 
%    \colhead{Nuclear Interaction} & 
    \colhead{$K$} & 
    \colhead{$m_n^*/m_n$} & 
    \colhead{$m_p^*/m_p$} & 
    \colhead{M$_{\mathrm{max}}$} & 
    \colhead{$R_{1.4 M_{\odot}}$} &
    \colhead{Refs.} \\
    \colhead{} &
    \colhead{} & 
%    \colhead{} & 
    \colhead{(MeV)} & 
    \colhead{} & 
    \colhead{} & 
    \colhead{(\msun)} & 
    \colhead{(km)} &
    \colhead{}
}
\startdata
1 & DD2   & 
    %DD2 & 
    $242.7$ & 0.5628 & 0.5622
    & $2.42$ & 13.2 & 1,2 \\
2 & SFHo   & 
    %SFH & 
    $245.4$ & 0.7609 & 0.7606
    & $2.06$ & 11.9 &3 \\
3 & SFHx   & 
    %SFH & 
    $238.8$ & 0.7179 & 0.7174
    & $2.13$ & 12.0 & 3 \\
4 & BHB$\lambda \phi$ & 
    %DD2, hyperons & 
    $242.7$  & 0.5628 & 0.5622
    & $2.10$ & 13.2 & 4 \\
5 & TM1    & 
    %TM1 & 
    $281.6$ & 0.6343 & 0.6338
    & $2.21$ & 14.5 & 1,2 \\
6 & NL3    & 
    %NL3 & 
    $271.5$ & 0.5954 & 0.5949
    & $2.79$ & 14.8 & 1,2 \\
\enddata
\tablecomments{
    Listed are the incompressibility $K$, the effective neutron and proton masses $m_n^*$ and $m_p^*$, the maximum mass of a cold neutron star, and the radius of a 1.4~\msun~neutron star. 
}
\tablerefs{
    (1)~ \cite{hempel2010}
    (2)~ \cite{hempel2012}
    (3)~\cite{Steiner.Hempel.Fischer.eos:2013}
    (4)~\citet{banik2014}
}
\end{deluxetable}

% \tablerefs{(1)~\citet{SFHo}
% (2)~\citet{DD2}
% (3)~\citet{hempel2012}
% (4)~\citet{banik2014}
% (5)~\citet{ls220}
% (6)~\citet{shen98_Thphy}
% (7)~\citet{shen98_nuphy}
% (8)~\citet{shen2011}
% }

\section{Estimation of $\lowercase{p_s}$}\label{app:ps}
The density $p_s$ is estimated with a weighted kernel density estimation procedure. Due to noise variable $(e)^s$, an additional Monte Carlo procedure is adopted, with the number of Monte Carlo samples $N_e=10,000$. Below is the corresponding pseudo code.

\begin{algorithm}[H]
\SetAlgoLined
\KwInput{$s$: equation of state; $N_e$: number of Monte Carlo samples}
\KwOutput{Estimate of $p_s$}
\For{$i=1,\dots, n_s$}{
Generate measurement errors $(e)^s_{i1},\dots,(e)^s_{iN_e}$ independently according to the density $p(e \mid x^s_i)$\;
Construct
\begin{align*}
    %(x)^s_{ij} = (x)^s_i
    (w)^s_{ij} = w_s((x)^s_{i})
    \quad \mbox{and}\quad
    (\by)^s_{ij} = (\bz)^s_{i} + (e)^s_{ij}, \quad j=1,\dots, N_e.
\end{align*}
}
Perform the weighted kernel density estimation on
the data $\{(\by)^s_{ij}: i=1,\dots, n_s; j=1,\dots, N_e\}$
with weights $\{(w)^s_{ij}\}$\;
\caption{Estimation of $p_s$}
\end{algorithm}

\section{Bootstrap Confidence Interval}
\label{app:bootstrap}

In order to obtain the confidence intervals of our metrics, we resort to a bootstrap procedure with the number of bootstrap samples $K=10,000$. The corresponding details are given in the following pseudo code.

\begin{algorithm}[H]
\SetAlgoLined
\KwInput{$s$: equation of state; $K$: number of bootstrap samples; $N_e$: number of Monte Carlo samples; $(1-\alpha)100\%$: confidence level}
\KwOutput{$(1-\alpha)100\%$ bootstrap confidence intervals of the estimated KL divergence and TV distance}
  \For{$j=1,\dots, K$}{
  Generate a bootstrap data set by sampling $n_s$ observations from $\{(x)_i^s, (\bz)_i^s\}_{i=1}^s$ with replacement\;
  Apply Algorithm 1 with the bootstrap data set (instead of the original data set $\{(x)_i^s, (\bz)_i^s\}_{i=1}^s$) and compute the KL divergence and the total variation distance, denoted by
  $(D_{KL})_j^s$ and $(D_{TV})_j^s$ respectively\;
  }
  Compute the $(\alpha/2)$-quantile and $(1-\alpha/2)$-quantile of $\{(D_{KL})_j^s\}_{j=1}^{K}$ to form a $(1-\alpha)100\%$ confidence interval for the estimated KL divergence\;
  Compute the $(\alpha/2)$-quantile and $(1-\alpha/2)$-quantile of $\{(D_{TV})_j^s\}_{j=1}^{K}$ to form a $(1-\alpha)100\%$ confidence interval for the estimated TV distance\;
 \caption{Bootstrap confidence interval}
\end{algorithm}

\end{document}